\documentclass[a4paper, 11pt]{article}

\usepackage{cite}
\usepackage{amsmath}
\usepackage{amssymb}
\usepackage{epsfig}
\usepackage{empheq}
\usepackage{subfigure,slashed}

\def\ba{\begin{eqnarray}}
\def\ea{\end{eqnarray}}
\def\be{\begin{equation}}
\def\ee{\end{equation}}

\def\nn{\nonumber}

\def\eps{\epsilon}

\def\R{{\mathcal R}}
\def\Q{{\mathcal Q}}

\def\F{{\mathcal F}}

\def\x{{\bf x}}
\def\y{{\bf y}}
\def\z{{\bf z}}
\def\k{{\bf k}}

\begin{document}

\setlength\arraycolsep{2pt}

\renewcommand{\theequation}{\arabic{section}.\arabic{equation}}
\setcounter{page}{1}

\begin{titlepage}

\begin{center}

\vskip 1.0 cm

{\LARGE  \bf Consistently violating the non-Gaussian consistency relation}

\vskip 1.0cm

{\Large
Sander Mooij and Gonzalo A. Palma
}

\vskip 0.5cm

{\it
\mbox{Grupo de Cosmolog\'ia y Astrof\'isica Te\'orica},
\mbox{Departamento de F\'{i}sica, FCFM, Universidad de Chile}\\ 
\mbox{Blanco Encalada 2008, Santiago, Chile}\\
}

\vskip 1.5cm

\end{center}

\begin{abstract} 

Non-attractor models of inflation are characterized by the super-horizon evolution of curvature perturbations, introducing a violation of the non-Gaussian consistency relation between the bispectrum's squeezed limit and the power spectrum's spectral index. In this work we show that the bispectrum's squeezed limit of non-attractor models continues to respect a relation dictated by the evolution of the background. We show how to derive this relation using only symmetry arguments, without ever needing to solve the equations of motion for the perturbations.

\end{abstract}

\end{titlepage}

\newpage

\setcounter{equation}{0}
\section{Introduction}

The measurement of departures from a purely Gaussian distribution of primordial curvature perturbations would give us access to exquisite details about the physics underlying cosmic inflation~\cite{Guth:1980zm, Linde:1981mu, Albrecht:1982wi}. Indeed, different models of inflation predict distinctive deviations from Gaussianity that are sensitive to the perturbations' self-interactions, as well as their interactions with other degrees of freedom that might have existed during inflation~\cite{Bartolo:2004if, Byrnes:2014pja}. This realization has motivated an extensive amount of research over the past decade on the subject of primordial non-Gaussianity, both theoretically~\cite{Allen:1987vq, Falk:1992sf, Gangui:1993tt, Acquaviva:2002ud, Maldacena:2002vr, Dvali:2003em, Zaldarriaga:2003my, Lyth:2005fi} and observationally~\cite{Senatore:2009gt, Liguori:2010hx, Komatsu:2010hc, Yadav:2010fz}. Although current constraints on non-Gaussianity coming from Cosmic Microwave Background (CMB) observations remain poor~\cite{Komatsu:2010fb, Ade:2013ydc, Ade:2015ava}, future Large Scale Structure surveys~\cite{Verde:2010wp, Desjacques:2010jw, Desjacques:2010nn} and Lyman-$\alpha$ Forest observations~\cite{Viel:2008jj, Hazra:2012qz, Seljak:2012tp} promise to substantially improve our knowledge about non-Gaussianity, giving us better insights into the elusive nature of inflation. 

Because a complete characterization of non-Gaussianity is somewhat out of reach, it is customary to parametrize non-Gaussian departures by defining the bispectrum $B_\R (\k_1, \k_2, \k_3)$, which determines the $3$-point correlation function of adiabatic curvature perturbations, hereby denoted by $\R$, in momentum space
\be
\langle \R (\k_1)   \R ( \k_2 ) \R (\k_3) \rangle \equiv (2 \pi)^3 \delta^{3} (\k_1+ \k_2 + \k_3)  B_\R (\k_1, \k_2, \k_3) ,
\ee
where the Dirac-delta function appears as a consequence of the homogeneity and isotropy of the inflationary background. At first order, the functional dependence of $B_\R$ on the three momenta $\k_1$, $\k_2$ and $\k_3$ is determined by the non-linear evolution of $\R$, parametrized by tree-level cubic interactions appearing in the Lagrangian describing its perturbative dynamics. As a consequence, in the simplest class of inflationary models ---namely single field slow-roll inflation--- one predicts a scale invariant bispectrum with an amplitude suppressed by the slow-roll parameters~\cite{Acquaviva:2002ud, Maldacena:2002vr}, but with a shape determined by the configuration of the momenta, restricted to the shell $\k_1+ \k_2 + \k_3=0$. More exotic single-field (non-canonical) models of inflation may predict substantially larger departures from Gaussianity~\cite{Seery:2005wm, Chen:2006nt}, some of them even showing strong departures from scale invariance~\cite{Adshead:2011jq, Behbahani:2012be, Gong:2014spa, Achucarro:2012fd, Palma:2014hra}. These models have been sys\-te\-ma\-ti\-cal\-ly analyzed with the help of the effective field theory of inflation approach~\cite{Cheung:2007st}, which allows one to study both canonical and non-canonical models of inflation within a general framework without the need of specifying the physics underlying inflation~\cite{Senatore:2010wk, Senatore:2009cf, Baumann:2011su, LopezNacir:2011kk, Achucarro:2012sm, Assassi:2013gxa}. In this framework, deviations from canonical inflation are parametrized by the sound speed at which curvature perturbations propagate (among other quantities). It is now well understood that a suppressed sound speed increases the strength of curvature perturbations' self-interactions, therefore enhancing the amount of primordial non-Gaussianity generated during inflation. Models characterized for allowing a suppressed value of the sound speed, with distinctive non-Gaussian shapes, include $P(X)$-inflation~\cite{ArmendarizPicon:1999rj, Seery:2005wm, Chen:2006nt}, DBI-inflation~\cite{Silverstein:2003hf, Alishahiha:2004eh}, and the low energy limit of multi-field inflation with heavy fields~\cite{Tolley:2009fg, Achucarro:2010da, Cespedes:2013rda}, just to mention a few. On the other hand, it is also possible to have large non-Gaussianity appearing as a consequence of nontrivial initial states~\cite{Holman:2007na, Meerburg:2009ys,Ashoorioon:2010xg,Ashoorioon:2013eia}.

One of the most outstanding discoveries in the study of non-Gaussianity is the so-called consistency relation, first reported by Maldacena in ref.~\cite{Maldacena:2002vr}. This relation tells us that in the squeezed limit (that is, the configuration where the size of one of the three momenta $\k_1$, $\k_2$ and $\k_3$ is much smaller than the remaining two) the bispectrum is determined by the power spectrum $P_{\R}$ and its spectral index $n_\R$ in the following specific way
\be
 \lim_{\k_3 \to 0} B_\R (\k_1, \k_2, \k_3) =  \big( n_\R(k_3) - 1 \big) P_{\R}(k_1)  P_{\R}(k_3) , \label{intro:squeezed-1} 
\ee
where $k_i \equiv |\k_i|$, and where $P_\R(k)$ and $n_\R$ are defined via:
\ba
 \langle \R(\k_1) \R(\k_2)\rangle&\equiv& (2\pi)^3 \delta^{(3)}\left(\k_1+ \k_2\right) P_\R(k_1) , \\ \qquad  n_\R(k)  &\equiv &1+ \frac{d }{d\ln k} \ln \left[ k^3  P_{\R}(k)  \right]. \label{powspecdef} 
\ea
Because both the amplitude of the power spectrum $P_{\R}$ and its spectral index $n_\R$ are fairly well measured~\cite{Ade:2013uln, Ade:2015oja}, this consistency relation is regarded as one of the most powerful tools available to falsify a large variety of single field inflationary models. It is not even restricted to slow-roll models of inflation\cite{Sreenath:2014nca}. Its elegant derivation, which will be reviewed in Section~\ref{consres}, combines simple symmetry arguments with simple statistical considerations that are independent of the dynamical details of inflation~\cite{Creminelli:2004yq, Cheung:2007sv, Creminelli:2011rh}. The basic assumption underlying its derivation is that, during inflation, the value of every background quantity (for example the Hubble parameter $H$) is uniquely determined by a single time-dependent parameter (for instance, the value of the inflaton vacuum expectation value, were we interested in the case of single field inflation). Models respecting this property are called attractor models, and they are further characterized by the fact that curvature perturbations freeze after horizon crossing. For this reason, a violation of the consistency relation, confirmed by observations, would automatically rule out every model of inflation in which the amplitude of curvature perturbations remained constant after horizon crossing, encompassing essentially all sensible single-field models of inflation.  

One of the preferred ways to constrain non-Gaussianity with the help of CMB observations is by defining the so called local $f_{\rm NL}$-parameter, which is related to the squeezed limit of the bispectrum through the relation
\be
 f_{\rm NL} \equiv -\frac{5}{12} \lim_{\k_3 \to 0} \frac{ B_\R (\k_1, \k_2, \k_3)}{ P_{\R}(k_1)  P_{\R}(k_3) } .
\ee
from where one reads $ f_{\rm NL} \simeq -5 (n_\R - 1) / 12 \simeq 0.02$ after using the latest constraints on the spectral index $n_\R$. However, projection effects in the measurement of the CMB preclude us from directly measuring this predicted value of $f_{\rm NL}$, implying that the consistency relation is in fact equivalent to $f_{\rm NL}^{\rm obs} = 0$~\cite{Pajer:2013ana}, which may be compared to the most recent constraints on local non-Gaussianity~\cite{Ade:2015ava}, given by $f_{\rm NL}^{\rm obs} = 0.8 \pm 5.0$. Thus, current observations are rather weak in providing a useful assessment of the validity of the consistency relation. Nevertheless, the prospects of measuring violations to the consistency relation~(\ref{intro:squeezed-1}) has motivated a fairly big amount of research devoted to the study of models of inflation where curvature perturbations are forced to evolve outside the horizon.\footnote{It is also possible to violate the consistency relation with a non-trivial initial state for curvature perturbations. See for instance refs.~\cite{Agullo:2010ws, Chialva:2011hc, Gong:2013yvl, Kundu:2013gha, Berezhiani:2014kga}.} The most prominent example of such models is multi-field inflation, where curvature perturbations have the chance to interact with other light degrees of freedom even after horizon crossing, allowing for large deviations of the consistency relation~\cite{Byrnes:2008wi, Byrnes:2008zy, Byrnes:2010em}. Another example is the case of warm inflation~\cite{Bastero-Gil:2014raa}, in which the inflaton remains coupled to a thermal bath. However, it was recently realized that curvature perturbations may also grow after horizon crossing in purely single field models characterized by a ``non-attractor" evolution of their background. For instance, the authors of ref.~\cite{Namjoo:2012aa} (see also~\cite{Martin:2012pe}) studied the generation of non-Gaussianity in a class of models known as ultra slow-roll inflation~\cite{Tsamis:2003px, Kinney:2005vj}, where the inflaton potential is exactly flat, precluding the existence of an attractor regime. In such a model, curvature perturbations grow outside the horizon at a dramatic rate, implying a squeezed limit for the bispectrum of the form
\be
 \lim_{\k_3 \to 0} B_\R (\k_1, \k_2, \k_3) = - 6 \, P_{\R}( k_1)  P_{\R}( k_3) , \label{intro:squeezed-2}
\ee
corresponding to $f_{\rm NL} = 5/2$. Given that in these models the power spectrum is almost scale invariant $n_\R \simeq 1$, we see that eq.~(\ref{intro:squeezed-2}) represents a flagrant violation of the consistency relation depicted in eq.~(\ref{intro:squeezed-1}). Furthermore, it has been argued that large violations of the consistency relation may constitute a generic feature of non-attractor models~\cite{Chen:2013eea, Akhshik:2015nfa} which may provide an alternative paradigm ---to that offered by slow-roll inflation--- in order to explain the generation and evolution of primordial curvature perturbations.

After accepting the fact that in this class of models the consistency relation is violated due to the super-horizon evolution of curvature perturbations, we ought to ask whether there are alternative explanations behind eq.~(\ref{intro:squeezed-2}) other than a brute force computation based on the dynamics of the perturbations. The purpose of this article is to clarify the underlying nature of the violation of the consistency relation in models with super-horizon evolution. To this extent, we will focus our discussion on single field models with non-attractor backgrounds. Our aim is to show that models displaying a violation of the consistency relation are still restricted to respect well defined relations between the squeezed limit of the bispectrum and the power spectrum, similar to Maldacena's consistency relation. We will show that, in fact, the same arguments leading to Maldacena's consistency relation, if phrased correctly, will still reproduce the result expressed in eq.~(\ref{intro:squeezed-2}) found in ref.~\cite{Namjoo:2012aa}. In other words, the squeezed limit may be determined purely from symmetry and statistical arguments, without the need of understanding the details about the dynamics of the specific system under interest. We will do this first for ultra slow-roll inflation, and then extend our results to more general non-attractor models. In this regard, we deduce a general expression determining the bispectrum's squeezed limit, given by
\be
 \lim_{\k_3 \to 0} B_\R (\k_1, \k_2, \k_3) = \frac{3}{c_s^2}(4+ \eta)P_{\R}( k_1)  P_{\R}( k_3),
\ee
where $\eta = \dot \epsilon / H \epsilon$ (with $\epsilon = - \dot H / H^2$ and $H$ being the Hubble expansion rate during inflation) and $c_s$ is the speed of sound of curvature perturbations. In this expression, the value of both $\eta$ and $c_s$ will depend on the specific model allowing for non-attractor solutions of the background.

We have organized this article as follows. In Section~\ref{consres} we review the derivation of the consistency relation~(\ref{intro:squeezed-1}) closely following the discussions found in refs.~\cite{Creminelli:2004yq, Cheung:2007sv}. Then, in Section~\ref{USR} we briefly review the model of ultra slow-roll inflation, which is the simplest model admitting a non-attractor behavior of the background. Section~\ref{suphor} is devoted to the study of the superhorizon behaviour of the inflaton and metric (scalar) perturbations in this model. In Section~\ref{comp} we show how the same arguments used to derive Maldacena's consistency relation can help to set up a modified relation valid for ultra slow-roll inflation, yielding the same result derived in ref.~\cite{Namjoo:2012aa} by a direct computation. Then, in Section~\ref{sec:general-non-attractor} we generalize the arguments developed in the previous sections to more general non-attractor models, based on $P(X)$-models of inflation. Finally, in Section~\ref{sec:conclusions}, we provide our concluding remarks. 

Before commencing, a quick word about units and notation: We shall use natural units whereby $c=1$, $\hbar = 1$ and $M_{\rm Pl}^2 = 1/8 \pi G_N = 1$. In addition, in this paper we formally denote the gauge invariant co-moving curvature perturbation by $\R$, which in co-moving gauge coincides with the spatial metric perturbation (usually denoted by $\psi$). Furthermore, there is a sign difference between our convention for $\R$ (and $\psi$) and the convention used in, among others, references \cite{Maldacena:2002vr, Namjoo:2012aa,Creminelli:2004yq}. See for example our metric in~(\ref{metric-co-moving}). At every stage where we cite results obtained in these references, we have accounted for that sign difference.

\setcounter{equation}{0}
\section{Review of the consistency relation} \label{consres}

In this section we offer a review of the derivation of the consistency relation~(\ref{intro:squeezed-1}), closely following the discussions of refs.~\cite{Creminelli:2004yq} and~\cite{Cheung:2007sv} (see also ref.~\cite{Ganc:2010ff, RenauxPetel:2010ty}). Let us start by recalling that the perturbed Friedman-Robertson-Walker (FRW) metric in co-moving gauge may be written with the help of the Arnowitt-Deser-Misner formalism~\cite{Arnowitt:1962hi} as
\be
ds^2 = - N^2 dt^2 + a^2(t) e^{- 2 \R } \delta_{i j} (d x^i + N^i dt)(d x^j + N^j dt) , \label{metric-co-moving}
\ee
where $a(t)$ is the scale factor parametrizing the expansion of spatial foliations during inflation, and $\R$ represents the adiabatic curvature perturbation. In addition, $N$ and $N^i$ are the usual lapse and shift functions respectively, to be determined in terms of other quantities by solving constraint equations. It is customary to define $\delta N$ through
\be
N \equiv 1 + \delta N,
\ee
in which case the background FRW metric is recovered by setting $\R = 0$, $\delta N = 0$ and $N^i = 0$. Let us notice that an immediate consequence of the metric~(\ref{metric-co-moving}) is a symmetry under simultaneous rescalements of the scale factor and the co-moving coordinates:
\be
a(t) \to a'(t) = a(t) e^{\Delta C} , \qquad dx \to dx' = dx e^{-\Delta C} . \label{FRW-symmetry}
\ee
This rescalement does not affect other observable background quantities, such as the Hubble expansion rate $H = \dot a / a$. In addition, notice that
the rescalement of the scale factor may be absorbed in the curvature perturbation 
\be
\R \to \R ' = \R - \Delta C,   \label{rshift}
\ee
implying that the equations of motion for $\R$ must admit at least one solution of the form $\R = {\rm constant}$, which reminds us that, after all, we can only measure gradients of $\R$.

\subsection{Attractor backgrounds and long wavelength modes} \label{attractor-long-wavelengths}

Our main focus in this section are models of inflation characterized by a background with an attractor behavior. These are models where every background quantity is uniquely determined by a single parameter, which may be used as a replacement of time. The simplest example of such backgrounds is offered by single field slow-roll inflation, where the background quickly asymptotes to an attractor trajectory respecting equations of motion of the form
\be
3 H \dot \phi + \frac{\partial V}{\partial \phi} = 0, \qquad 3 H^2 = V(\phi) ,
\ee
where $H= \dot a / a$ is the Hubble expansion rate. These equations tell us that both $H$ and $\dot \phi$ are completely determined by the value of $\phi$, irrespective of the initial conditions. Universes following this feature have been dubbed single-clock universes by the authors of ref.~\cite{Creminelli:2004yq}. There is a direct consequence on the dynamics of perturbations springing out from this behavior. First, notice that since we are working in co-moving gauge (that is, scalar perturbations are only present in the metric via eq.~(\ref{metric-co-moving})), then every patch of the universe is determined by the same value of $\phi$, implying that every patch is characterized by the same Hubble pa\-ra\-me\-ter $H$. Second, recall that the wavelength of any perturbation in a FRW background grows proportionally to the scale factor $a(t)$. This means that we can split both $\R$ and $\delta N$ in short and long wavelength contributions
\be
\R = \R_s + \R_\ell , 	\qquad  \delta N = \delta N_s + \delta N_\ell , \label{split-short-long}
\ee
where $\R_\ell$ and $\delta N_\ell$ contain contributions with frequencies smaller than $H$. A physical observer who has access only to short wavelength perturbations will not be able to distinguish the long wavelength contributions $\R_\ell$ and $\delta N_{\ell}$ from the background. To understand the consequence of this, let us consider a reference time $t_0$ and insert back the long wavelength contributions of the splitting~(\ref{split-short-long}) in the metric~(\ref{metric-co-moving}). We find
\be
ds^2 \big |_{\x} = -  dt_B^2 + a_{\rm eff}^2(t_B,\x) \delta_{i j} d x^i d x^j  , \label{eff-metric}
\ee
where $t_B$ and $a_{\rm eff}$ are given by:
\ba
t_B (t,  \x) &=& t +  \delta t (t,  \x), \qquad \delta t (t,  \x) \equiv \int^t_{t_0} \!\! dt'  \delta N_{\ell} (t' , \x) , \label{t-B} \\
a_{\rm eff} (t_B,\x) &=& a( t(t_B) ) e^{-\R_\ell (t, \x)} , \label{a-eff}  \\
\phi_{\rm eff} (t_B,\x) &=& \phi (t(t_B)) ,  \label{phi-eff}
\ea
where $t(t_B) = t_B - \delta t$ (notice that we have conveniently set $\delta t = 0$ for $t = t_0$). Now, we wish to examine how the long wavelength fields affect the local background about $\x$ at the vicinity of $t=t_0$. First, notice that $a_{\rm eff}$ has to be a solution of the same equations of motions respected by the original solution $a(t)$. Given that proper time at the $\x$-patch is given by $t_B$, the only possibility for $a_{\rm eff}$ to satisfy these equations is that it consists of $a$ evaluated at $t_B$, up to a multiplicative constant allowed by the symmetry (\ref{FRW-symmetry}) of the background:
\be
a_{\rm eff}  = a(t_B) e^{- \Delta C} . \label{a_eff-a_tB}
\ee
By equating this with (\ref{a-eff}) we find 
\be
a(t) e^{-\R_{\ell}} = a(t + \delta t) e^{- \Delta C} . \label{a-delta-t}
\ee
Then, expanding the right hand side about $t$, we obtain
\be
a(t) e^{-\R_{\ell}} \simeq a(t ) \left[ 1 + H \delta t(t,  \x) \right] e^{- \Delta C} \simeq a(t) e^{ H \delta t (t,\x)- \Delta C} . \label{a-eff-delta-t}
\ee
Comparing this result with~(\ref{a_eff-a_tB}) we see that $\R_\ell =  \Delta C - H \delta t $, from where we derive that $\dot \R_\ell = - \dot H \delta t - H \delta N_\ell $. Then, by disregarding $\dot H \delta t \simeq \dot H N_\ell (t - t_0)$, which is sub-leading compared to the other two terms, we conclude that:
\be
\delta N_\ell = - \frac{\dot \R_\ell}{H} .
\ee
Notice that this is the usual relation obtained by solving the constraint equation for the shift $\delta N$ at linear order. While this result is valid for all wavelengths, our derivation is only valid for long wavelength modes.

Up to this point our arguments have been rather general, and valid for both attractor and non-attractor backgrounds. To see the implications of dealing with an attractor background, let us go back to eq.~(\ref{phi-eff}). Recall that, since we are in co-moving gauge, in an attractor background there is only one value of $\phi$ characterizing the background at a given time $t$. In particular, the value of the background field at the local patch centered at $\x$ must coincide with $\phi(t_B)$:
\be
\phi_{\rm eff} (t_B,\x)   = \phi (t_B).
\ee
By comparing this expression with (\ref{phi-eff}) we see that $\delta N_\ell = 0$. In addition, it also implies that 
\be
\R_{\ell} = {\rm constant} ,
\ee
is the only allowed behavior for long wavelength curvature perturbations in attractor backgrounds. In other words, as their wavelengths are stretched, curvature perturbations must freeze (implying that the second long wavelength mode must decay).

\subsection{The consistency relation} \label{consistency-computation}

Having established some properties of attractor backgrounds, let us now proceed to derive the consistency relation. We want to compute $\langle \R(\k_1) \R(\k_2) \R(\k_3)\rangle$ in the squeezed limit where $k_3\ll k_1,k_2$. In this limit, the long mode $k_3$ has left the horizon much earlier than those parametrized by $k_1$ and $k_2$. We can then compute the correlation function between $\R(\k_1)$ and $\R(\k_2)$ in a background renormalized by $\R(\k_3)$. We begin in position space (parametrized by co-moving coordinates) by computing the two-point correlation function $\langle \R \R \rangle(\x_1, \x_2) \equiv \langle \R(\x_1)\R(\x_2) \rangle$ in a patch of the universe centered at $(\x_1+\x_2)/2$ and inflating according to the effective scale factor:
\be
a_{\rm eff} = a (t) e^{- \bar \R_{\ell}} ,  \qquad  \bar \R_\ell = \R_\ell \Big(\frac{\x_1+\x_2}{2}  \Big) .  \label{xrescal}
\ee
Now, we may choose to rescale $a_{\rm eff}$ back to $a$ as long as we properly rescale the co-moving coordinates. In other words, we may write
\be
\langle \R \R \rangle_B(\x_1, \x_2) = \langle \R \R \rangle_0( e^{- \bar  \R_\ell} \x_1, e^{- \bar  \R_\ell}  \x_2) ,
\ee
where $\langle \R \R \rangle_0$ denotes the computation of the two-point correlation function in a background with an expansion dictated by $a(t)$ alone.  However, since $\langle \R \R \rangle_0 (\x_1, \x_2) $ is just the two-point correlation function computed in the true homogenous and isotropic background, the result must depend on the difference $|\x_1 - \x_2|$. This further implies that:
\be
\langle \R \R \rangle_B(\x_1, \x_2) = \langle \R \R \rangle_0( e^{- \bar \R_\ell} | \x_1- \x_2| ).
\ee
We may now Taylor expand about $\R_\ell = 0$. Keeping the first two terms of the expansion, it is straightforward to find:
\be
\langle \R \R \rangle_B(\x_1, \x_2) =  \langle \R \R  \rangle_0 (| \x_1-\x_2|) - \R_\ell \Big(\frac{\x_1+\x_2}{2}  \Big) \frac{d}{d\ln{| \x_1-\x_2|}}  \langle\R \R \rangle_0  ( | \x_1-\x_2| ) + \cdots  .
\ee
Next, we can move the expressions to Fourier space. By performing a Fourier transformation with respect to $\x_1$ and $\x_2$, it is direct to find
\be
 \langle\R \R \rangle_B(\k_1, \k_2) \simeq  \langle \R  \R  \rangle_0 ( k_s) + \R_{\ell} \left(\k_\ell  \right) \frac{1}{k_s^3} \frac{d}{ d \ln{k_s}} \Big[k_s ^3 P_\R  ( k_s) \Big], \label{answer}
\ee
where $\k_\ell \equiv \k_1+\k_2$ and $\k_s \equiv (\k_1-\k_2)/2$.  As a last step, we may correlate the result of eq.~(\ref{answer}) with $\R(\k_3)$. By doing this, the first term at the right hand side of eq.~(\ref{answer}) averages out, returning:
\be
 \langle \langle\R \R \rangle_B(\k_1, \k_2)   \R (\k_3) \rangle  \simeq \langle \R_{\ell} (\k_\ell  ) \R (\k_3) \rangle  \frac{1}{k_s^3} \frac{d}{ d \ln{k_s}} \Big[ k_s ^3 P_\R  ( k_s) \Big]. \label{answer-2}
\ee
The left hand side of this expression gives us back $ \langle \R (\k_1) \R (\k_2) \R (\k_3)  \rangle_{\k_3 \to 0}$, finally leading to the desired result
\be
 \lim_{\k_3 \to 0} B_\R (\k_1, \k_2, \k_3) = \big( n_\R(k_3) - 1 \big) P_{\R}( k_1)  P_{\R}( k_3)  , \label{consrel}
\ee
where the spectral index is identified with:
\be
n_\R (k)  = 1 + \frac{d }{d \ln{k}}  \ln \left[ k^3 P_\R(k) \right] .
\ee
The consistency relation tells us that any measurement of $f_{\rm NL}$ larger than $1-n_\R \simeq 0.04$ will rule out every model in which the superhorizon perturbations $R_\ell$ freeze, i.e. every attractor adiabatic single field slow-roll inflation model. On the other hand, in the next section we will see that in the non-attractor model of ultra slow-roll inflation, the modes $\R_\ell$ evolve rapidly outside the horizon. Then it does not come as a surprise that the consistency relation breaks down, as reported in \cite{Namjoo:2012aa}.

\setcounter{equation}{0}
\section{Non-attractor backgrounds: Ultra slow-roll inflation} \label{USR}

In this section we examine some aspects of non-attractor models, where the evolution of the background keeps some memory of the initial conditions. To keep our discussion simple we will focus our attention on the particular case of ultra slow-roll inflation studied in refs.~\cite{Tsamis:2003px, Kinney:2005vj}. We shall examine more general non-attractor backgrounds in Section~\ref{sec:general-non-attractor}. 

\subsection{Ultra slow-roll inflation}

Ultra slow-roll inflation was first introduced in ref.~\cite{Tsamis:2003px} and further worked out in ref.~\cite{Kinney:2005vj}. In its simplest version, ultra slow-roll inflation is realized by an exactly flat scalar field potential of the form
\be
V(\phi) =V_0,  \label{flat-potential}
\ee
where $V_0$ is a constant. Evidently, this potential does not constitute a very realistic choice (see for example the discussion in \cite{Martin:2012pe}) but it allows us to study the evolution of perturbations in backgrounds that are dramatically different from those encountered in conventional slow-roll inflation. Let us notice that eq.~(\ref{flat-potential}) automatically implies that the theory is invariant under the shift symmetry $\phi \to \phi ' = \phi + \Delta \phi$ in addition to the rescaling of the scale factor examined in Section~\ref{consres}. We will come back to the consequences of this symmetry in a moment. The background equations of motion of the system are 
\ba
\ddot \phi + 3 H \dot \phi = 0 , \\ 
6 H^2 =  \dot \phi^2 + 2 V_0 .
\ea
The first equation already tells us that there is no such thing as an attractor behavior dominated by the friction term $3 H \dot \phi$ except for the trivial solution $\dot \phi = 0$. By adopting the notation $H= H(\phi)$, these equations can be combined into a single equation given by:
\be
3 H^2 - 2 ( H' )^2 = V_0 .
\ee
The solution to this second equation gives us $H$ in terms of $\phi$ once we have adopted a set of boundary conditions. For definiteness, we choose these conditions in such a way that $3 H^2 = V_0$ at $\phi = 0$. This condition implies:
\be
H(\phi) = \sqrt{\frac{V_0}{3}} \cosh \left( \sqrt{3/2}  \, \phi \right) . \label{solh}
\ee
To continue, we may choose to study inflation in the range $\phi <  0$. Then, the equation of motion for $\phi(t)$ may be integrated once to give:
\be
\dot \phi = - \sqrt{2V_0} \sinh \left(  \sqrt{3/2}  \, \phi \right) .  \label{dphi}
\ee
These equations permit us to compute various background quantities in terms of $\phi$. For instance, the scale factor $a(\phi)$ is found to be 
\be
a(\phi) = a_0 \left[ - \sinh  \left(  \sqrt{3/2}  \, \phi \right) \right]^{-1/3} ,
\ee
whereas the slow-roll parameters $\epsilon\equiv -\dot{H}/H^2$, $\eta \equiv \dot \epsilon / ( \epsilon H)$ and $\xi \equiv \dot{\eta}/(\eta H)$ are respectively found to be given by
\ba
&& \epsilon (\phi) = 3 \tanh^2 \left( \sqrt{3/2}  \, \phi \right) , \\
&& \eta (\phi) =  - 6 \cosh^{-2} \left( \sqrt{3/2}  \, \phi  \right) ,\\
&& \xi (\phi) = 6 \tanh^2 \left( \sqrt{3/2}  \, \phi \right) .
\ea
These results, in combination with eq.~(\ref{dphi}), show that in the limit $t \to + \infty$ the background parameters have the following asymptotic behavior: 
\be
 \epsilon \to a^{-6} , \qquad \eta \to -6, \qquad \xi \to a^{-6} .
\ee
Thus we see that this inflationary background is characterized by a substantially large value of $\eta$, offering a large departure from conventional slow-roll inflationary models. In what follows, we will restrict our analysis to the particular case of $\epsilon \ll 1$, which is necessary to reproduce a scale invariant power spectrum (See appendix~\ref{app:usr}).

\subsection{Long wavelength perturbations} \label{sec:long-wavelengths-flat}

Just as we did in Section~\ref{attractor-long-wavelengths}, we may infer the behavior of long wavelength modes in co-moving gauge by examining the symmetries of the non-attractor background at hand. To start with, recall that the ultra slow-roll background of eq.~(\ref{flat-potential}) is characterized by the symmetry:
\be
\phi \to \phi' = \phi + \Delta \phi . \label{symm-usr}
\ee
Let us emphasize that, as opposed to the case of attractor backgrounds, in ultra slow-roll inflation there are no solutions uniquely linking (background quantities like) $H$ and $\phi$. A variation of the initial conditions leads to a variation of the relation between $H$ and $\phi$. This implies that, as the wavelength of perturbations are stretched by the expansion of space, the long wavelength contribution to $N_\ell$ of eq.~(\ref{split-short-long}) may modify the background value of $\phi$ felt by sub-horizon modes, just as in eq~(\ref{phi-eff}), in a way consistent with the symmetry~(\ref{symm-usr}). In other words, this time, in addition to eq.~(\ref{a_eff-a_tB}), we may have
\be
\phi_{\rm eff} (t_B , \x) = \phi (t_B) + \Delta \phi , \label{phi-eff-C}
\ee
where $\Delta \phi$ is a constant incorporating the effects of long-wavelength perturbations via the shift function $N_{\ell}$ (but should not be confused with perturbations of the field $\phi$, which in co-moving gauge are turned off). From (\ref{phi-eff}) we see that the previous equation is equivalent to
\be
\phi (t) = \phi (t + \delta t) + \Delta \phi . \label{phi-delta-t}
\ee
This implies that to linear order:
\be
\delta t = - \frac{\Delta \phi}{\dot \phi} .
\ee
Then, using this result in combination with (\ref{a_eff-a_tB}) and (\ref{a-eff}) we obtain
\be
\R_\ell = \Delta C + \frac{H}{\dot \phi} \Delta \phi .  \label{R_ell_C2}
\ee
Recalling eqs.~(\ref{solh}) and (\ref{dphi}) of the previous section, we see that the asymptotic behavior of $\R_\ell$ soon or later will be dominated by $\Delta \phi$ which implies a growing mode of the form:
\be
\R_\ell \sim  a^3 .  \label{growmod}
\ee
That is, ultra slow-roll inevitably contains super-horizon evolution of curvature perturbations. We are for sure not the first ones to achieve this result, see for example \cite{Kinney:2005vj,Namjoo:2012aa}. However, we would like to emphasize  that in this case it has been exclusively deduced with help of symmetry considerations.

\subsection{A violation of the consistency relation?} \label{consvio}

The result expressed in eq.~(\ref{growmod}) tells us that the argument used to derive the consistency relation involving the rescalement of the co-moving coordinates cannot be repeated without carefully taking into account the additional contribution from $\phi_{\rm eff}$ felt by sub-horizon modes. Indeed, a brute force computation of the squeezed limit of non-Gaussianity leads to a violation of the consistency relation~\cite{Namjoo:2012aa} in the form:
\be
 \lim_{\k_3 \to 0} B_\R (\k_1, \k_2, \k_3)  =  - 6 P_{\R}(\k_1)  P_{\R}(\k_3)   . \label{cr-1}
\ee
This result signals a violation of the consistency relation~(\ref{consrel}) which in slow-roll attractor backgrounds prescribes
\be
 \lim_{\k_3 \to 0} B_\R (\k_1, \k_2, \k_3) = - ( \eta + 2 \epsilon ) P_{\R}(\k_1)  P_{\R}(\k_3) ,\label{cr-2}
\ee
where we have used the standard relation between the spectral index and the slow-roll parameters $1 - n_\R  = \eta + 2 \epsilon$.
In this case, the mere fact that a model of adiabatic single field slow roll inflation can produce an order-one non-Gaussianity (one gets $f_{\rm NL} = 5/2$) shows the apparent breakdown of the consistency condition\footnote{Note that the disagreement between (\ref{cr-1}) and (\ref{cr-2}) continues to hold in the decoupling limit: inserting $\eps \to 0$ and $\eta\to -6$ still produces a sign difference between the two. It is of course not completely fair to compare the two results in such a way, since the derivation of the result~(\ref{cr-2}) supposes $\eps,\eta\ll1$.}. Thus, as the consistency relation is supposed to be valid for all adiabatic single-clock models of inflation, the model of ultra slow-roll inflation seems to provide a counterexample.  However, in section \ref{comp} we will show how a generalized version of the consistency relation still holds for the model of ultra slow-roll inflation.

\setcounter{equation}{0}
\section{Freezing of superhorizon perturbations} \label{suphor}

Before analyzing the squeezed limit in the context of non-attractor models, let us briefly analyze the non-linear relation between curvature perturbations and inflaton perturbations in the context of ultra slow-roll. Up to this point, we have been working with adiabatic perturbations $\R$ in co-moving gauge, where inflaton perturbations are absent. More generally, $\R$ is a gauge invariant quantity that reduces to spatial curvature perturbations of the metric $\psi$ under the condition that inflaton perturbations $\delta \phi$ vanish:
\be
\R |_{\delta \phi =0} = \psi .
\ee
Alternatively, we may work in flat gauge, and define a gauge invariant perturbation $\Q$ that reduces to inflaton perturbations $\delta \phi$ under the condition that spatial curvature perturbations of the metric $\psi$ vanish:
\be
\Q |_{\psi=0}=\delta \phi,
\ee
It is then possible to find (see for instance~\cite{Maldacena:2002vr}) that $\R$ and $\Q$ are non-linearly related, up to quadratic order, in the following way:
\be
\R =  \frac{H}{\dot{\phi}} \Q + \frac{\eta}{4} \frac{H^2}{\dot{\phi}^2} \Q^2 -\frac{H}{\dot{\phi}^2} \Q \dot{\Q}  + \cdots , \label{R-Q}
\ee
or equivalently
\be
\Q = \frac{\dot{\phi}}{H}\R +\frac{\dot{\phi}}{H} \frac{\eta}{4}\R^2  +\frac{\dot{\phi}}{H^2}\R \dot{\R} + \cdots  ,  \label{Phirew}
\ee
where the ellipses ``$\cdots$'' represent terms with spatial gradients. Now, as we have seen, in conventional slow-roll inflation $\R$ freezes at long wavelengths, implying that $\dot \R_\ell \to 0$ fast enough to imply the relation:
\be
\textrm{ Slow-roll: } \qquad  \Q_\ell  = \frac{\dot{\phi}}{H}\R_\ell +\frac{\dot{\phi}}{H} \frac{\eta}{4}\R_\ell^2 , \qquad \R_\ell  = \frac{H}{\dot{\phi}} \Q_\ell  - \frac{H^2}{\dot{\phi}^2} \frac{\eta}{4}\Q_\ell^2 .
\ee
However, in the case of ultra slow-roll we have that $\R_\ell \propto a^3$, from where we read $\dot \R_\ell = 3 H \R_\ell$, giving us back the relation
\be
\textrm{ Ultra slow-roll: } \qquad  \Q_\ell  = \frac{\dot{\phi}}{H}\R_\ell - \frac{\dot{\phi}}{H} \frac{\eta}{4}\R_\ell^2 ,  \qquad \R_\ell  = \frac{H}{\dot{\phi}} \Q_\ell  + \frac{H^2}{\dot{\phi}^2} \frac{\eta}{4}\Q_\ell^2 ,  \label{QR-usr}
\ee
where we have used $\eta = - 6 + 2 \epsilon$. To exploit these relations, let us examine what happens if we were to expand a few quantities encountered in Section~\ref{consres} up to second order in the long wavelength perturbations. First, if we expand (\ref{a-delta-t}) about $t$ up to second order we find:
\be
a(t) e^{-\R_{\ell}} \simeq a(t) \left[ 1 + H \delta t(t,  \x) + \frac{1}{2} H^2 (1 - \epsilon) \delta t^2 \right] e^{- \Delta C} \simeq a(t) e^{ H \delta t (t,\x) - \frac{1}{2} \epsilon H^2 \delta t^2 - \Delta C } . \label{a-eff-delta-t-2}
\ee
This relation tells us that 
\be
\R_{\ell} = \Delta C -  H \delta t (t,\x) + \frac{1}{2} \epsilon H^2 \delta t^2 . \label{R-ell-delta-t}
\ee
On the other hand, by expanding (\ref{phi-delta-t}) about $t$ to second order, we find
\ba
\phi (t) &\simeq& \phi (t) + \dot \phi \delta t + \frac{1}{2} \ddot \phi \delta t^2 + \Delta \phi \nn \\
&\simeq& \phi (t) + \dot \phi \delta t -  \frac{3}{2} H \dot \phi \delta t^2 + \Delta \phi , \label{phi-delta-t-2}
\ea
where we used the background equation of motion $\ddot \phi + 3 H \dot \phi = 0$. This equation allows us to deduce $\delta t$ up to second order in $\Delta \phi$:
\be
 \delta t  = - \frac{\Delta \phi}{\dot \phi} + \frac{3}{2} \frac{H}{\dot \phi^2} \Delta \phi^2 .
\ee
Plugging this result back into eq.~(\ref{R-ell-delta-t}) we finally deduce
\be
\R_\ell  =  \Delta C +  \frac{H }{\dot \phi} \Delta \phi - \frac{3 - \epsilon}{2} \frac{H^2}{\dot \phi^2} \Delta \phi^2 ,
\ee
which, after noticing that $\eta = -6 + 2 \epsilon$, we see that is precisely consistent with (\ref{QR-usr}) once we replace $\Q_{\ell} \to \Delta \phi$ (and disregard $\Delta C$). Thus, we see that $\Q$ must freeze in order to have a long wavelength limit consistent with the symmetries of the background. In fact, a brute force computation reassures us that $\Q$ indeed freezes~\cite{Kinney:2005vj}. For instance, to second order, the action for $\Q$ is found to be given by:
\be
S_\Q^{(2)}=\frac{1}{2}\int d^4 x a^3 \left[ \dot{\Q}^2-2\eps V_0 \Q^2\right] . \label{sphi}
\ee
Notice the presence of the mass term proportional to $\epsilon$ which in fact breaks the shift symmetry $\phi \to \phi' = \phi + \Delta \phi$ shared by the background. However, since $\epsilon \ll 1$ and $\epsilon \propto a^{-6}$ it is still compatible with the freezing of $\Q$. To see this, 
we may perform the following field redefinition $u=\Q / f(\phi)$ with $f$ a function of the background field $\phi$. The quadratic action in terms of $u$ then reads
\be
S_u^{(2)}=\frac{1}{2}\int d^4 x a^3\left( f^2\dot{u}^2+ \dot \phi^2 \left[f f''+\cdots  \right]u^2\right) ,
\ee
where the ellipses ``$\cdots $" represents other terms containing $f$ and its derivatives with respect to $\phi$.  To find the freezing solutions for $u$, the part in square brackets above needs to vanish. This gives us a second order equation for $f$. One of the solutions will correspond to $u=\R$ (there is still a constant mode in $\R$, but it is subdominant compared to the other mode which grows as $a^3$).  The other solution corresponds to $u=\Q(1+\mathcal{O}(\eps))$, which is just another way of stating that (the dominant mode of) $\Q$ freezes up to all orders in the perturbations in the decoupling limit $\eps \to 0$.

\setcounter{equation}{0}
\section{A consistency relation for ultra slow-roll inflation}  \label{comp}

Our next goal is to compute $\langle \R(\k_1) \R(\k_2) \R(\k_3)\rangle$ in the squeezed limit $k_3\ll k_1,k_2$ for ultra slow-roll inflation, employing symmetry arguments as in the case of attractor models. As we have discussed in the previous sections, the main challenge is to take into account the super-horizon evolution of curvature perturbations $\R$. To proceed, we adopt the flat gauge, whereby spatial curvature perturbations $\psi$ are turned off. In this gauge the metric line element takes the form
\be
ds^2 = - N^2 dt^2 + a^2(t) \delta_{i j} (d x^i + N^i dt)(d x^j + N^j dt) , \label{metric-flat}
\ee
and scalar perturbations enter as excitations of the inflaton field:
\be
\phi (\x , t) = \phi(t) + \Q (\x , t)  .
\ee
As argued in the previous section, the symmetry of the theory implies that $\Q (\x , t)$ asymptotes to a constant on super-horizon scales. However, given that we derived this result in co-moving gauge, it is instructive to show how this result is recovered in flat gauge. This time, we must split both $\Q$ and $N$ into short and long wavelength contributions of the form
\be
\Q = \Q_s + \Q_\ell , \qquad N = N_s + N_\ell . \label{split-short-long-flat}
\ee
Then, a patch of the universe centered at $\x$ will be characterized by a background with an effective field $\phi_{\rm eff}$ and an effective scale factor $a_{\rm eff}$ given by
\ba
\phi_{\rm eff} (t_B,\x) &=& \phi ( t(t_B) ) + \Q_\ell (t, \x),  \label{phi-eff-flat} \\
a_{\rm eff} (t_B,\x) &=& a(t(t_B)) , \label{a-eff-flat}
\ea
where $t(t_B) = t_{B} - \delta t$ and $t_B$ is given by eq.~(\ref{t-B}). Now, notice that the only form for $a_{\rm eff}$ consistent with the symmetries of the background is $a_{\rm eff} = a (t_B) e^{\Delta C}$, with $\Delta C$ a constant.  This implies that:
\be
a(t) = a(t + \delta t) e^{- \Delta C} \simeq a e^{- \Delta C+ H \delta t} .
\ee
Then, we see that $\delta t = \Delta C / H$, or equivalently, $\delta N_\ell \propto d H^{-1} / dt$. As a consequence, $\delta N_\ell$ vanishes quickly
\be
\delta N_\ell \propto \epsilon \to a^{-6} ,
\ee
and the only long wavelength contribution to $\phi_{\rm eff}$ is via $\Q_\ell$ as:
\be
\phi_{\rm eff} (t, \x) = \phi(t) + \Q_\ell (t, \x) .
\ee
To continue, given that $\Q_\ell \to {\rm constant}$, we may compute the three-point correlation function $\langle \Q(\k_1) \Q(\k_2) \Q(\k_3)\rangle$ in the squeezed limit $k_3\ll k_1,k_2$ using the same arguments of Section~\ref{consistency-computation}. To this extent, we start by computing the two-point correlation function $\langle \Q \Q \rangle(\x_1, \x_2) \equiv \langle \Q(\x_1)\Q(\x_2) \rangle$ in a patch of the universe, centered at $(\x_1+\x_2)/2$ which feels a background given by
\be
\phi_{\rm eff} = \phi (t) + \Q_\ell \Big(\frac{\x_1+\x_2}{2}  \Big)  .
\ee
The two-point correlation function $\langle \Q \Q \rangle_B(\x_1, \x_2)$ computed in this background may then be Taylor expanded about the background without the long wavelength contributions, giving us back
\ba
\langle \Q \Q \rangle_B(\x_1, \x_2)& \simeq &  \langle\Q \Q \rangle_0 (| \x_1-\x_2|) + \Q_\ell \Big(\frac{\x_1+\x_2}{2}  \Big) \Bigl[\frac{d}{ d \Q_\ell} \langle\Q \Q \rangle_B ({\x_1}, {\x_2})  \Bigr]_0 \nn \\
& \simeq &  \langle\Q \Q \rangle_0 (| \x_1-\x_2|) + \Q_\ell \Big(\frac{\x_1+\x_2}{2}  \Big) \frac{d}{ d \phi} \langle\Q \Q \rangle_0 (\x_1, \x_2) ,
\ea
where we have traded the derivative with respect to $\Q_\ell$ with a derivative made with respect to $\phi$, the background value of the scalar field.
The next steps are straightforward. First, we move to Fourier space by performing a transformation with respect to $\x_1$ and $\x_2$ to obtain
\be
 \langle\Q \Q \rangle_B(\k_1, \k_2) \simeq  \langle \Q  \Q  \rangle_0 ( k_s) + \Q_{\ell} \left(\k_\ell  \right)  \frac{\partial}{\partial \phi}  P_{\Q}( k_s) ,   \label{answer-phi}
\ee
where $\k_\ell \equiv \k_1+\k_2$ and $k_s \equiv (\k_1-\k_2)/2$. Then we correlate the result with a mode $\Q (\k_3 )$. This gives us the squeezed limit:
\be
 \lim_{\k_3 \to 0} \langle \Q (\k_1) \Q (\k_2) \Q (\k_3 )  \rangle  = (2 \pi)^3  \delta^3 (\k_1 + \k_2 + \k_3) P_{\Q}( k_1) \frac{\partial}{\partial \phi}  P_{\Q}(k_3) .
\ee
To obtain a more explicit result we need an expression for the power spectrum $P_{\Q}(k)$ in terms of the background quantities.\footnote{Notice that we only need the amplitude of the power spectrum in terms of the background fields evaluated at a pivot scale, and we need not to worry about its scale dependence.} To this extent, it suffice to know that the power spectrum is proportional to $H^2$ (see for example~\cite{Kinney:2005vj}) 
\be
P_{\Q}(k) \propto H^2 .
\ee
Then, knowing that $H = H(\phi)$, we may write $\partial_\phi P_{\Q}(k) = 2 P_{\Q}(k) \partial_\phi H / H$, which together with the fact that $\partial_\phi H / H = \sqrt{\epsilon /2}$ (easily derived from eq.~(\ref{solh})) leads to
\be
 \lim_{\k_3 \to 0} \langle \Q (\k_1) \Q (\k_2) \Q (\k_3)  \rangle  = \sqrt{2\epsilon } \,(2 \pi)^3  \delta^3 (\k_1 + \k_2 + \k_3) P_{\Q}(\k_1)  P_{\Q}(\k_3) \label{epseq} .
\ee
We can now use this result to find the three-point function of $\R$. In eq.~(\ref{Phirew}) we already found how to rewrite $\Q$ in terms of $\R$ and $\dot{R}$ to quadratic order. Using that in ultra slow-roll we have $\R \sim a^3$ while $\eta\simeq -6$, then we can write
\be
\Q = \frac{\dot{\phi}}{H}\left[ \R -\frac{\eta}{4} \R^2    \right] + \cdots .  \label{crucrel}
\ee
where the ellipses ``$\cdots$'' stand  for spatial gradients, which in the present discussion may be disregarded. Note that the $\dot{\R}$ term effectively changes the sign of the correction to the first order result. This is to be equivalent to the observation made in~\cite{Namjoo:2012aa} that neglecting the contributions of the decaying mode (i.e. the $\dot{R}$-effects) changes the sign of the final result for the computation of $f_{\rm NL}$. To continue, notice that in general, for a relation of the form $f=g+\lambda g^2$ (see for example \cite{Maldacena:2002vr}) one finds:
\ba
 \langle f (\x) f (\y) f (\z) \rangle
&=&  \langle g (\x)  g (\y)  g (\z) \rangle +2\lambda \left[  \langle g (\x)g (\y) Ê\rangle  \langle g (\x)  g (\z) \rangle  + {\rm cyclic} \right]. \label{3-point-f-g}
\ea
Then, by choosing $f \equiv  \Q / \sqrt{2 \epsilon} $, $g \equiv \R$ and $\lambda \equiv - \eta /4$, we obtain
\be
 \langle \mathcal R (\x) \mathcal R (\y) \mathcal R (\z) \rangle  =  \left(2 \epsilon \right)^{-3/2} \langle \Q (\x) \Q (\y) \Q (\z) \rangle  +\frac{\eta}{2} \left[  \langle \mathcal R (\x) \mathcal R (\y) Ê\rangle  \langle  \mathcal R (\x)  \mathcal R (\z) \rangle  + {\rm cyclic} \right] .
\ee
To finish, we just need Fourier transform this expression, take the squeezed limit and replace the three-point correlation function by our previous result~(\ref{epseq}), to obtain:
\be
 \lim_{\k_3 \to 0}\langle \R(\k_1)\R(\k_2)\R(\k_3)\rangle = (2\eps+\eta) ~(2\pi)^3\delta^3 (\k_1 + \k_2 + \k_3) P_{\R}(\k_1)  P_{\R}(\k_3).  \label{newrel}
\ee
This coincides with the result~(\ref{cr-1}) obtained in \cite{Namjoo:2012aa} in the decoupling limit $\eps\to 0$ and $\eta\to -6$. We wish to stress, once more, that to derive this result we never needed to deduce the explicit solution of $\R$ nor $\Q$. Instead, we have only used symmetry arguments to relate long wavelength modes to the background. It is true, however, that even without explicitly deriving it, to deduce our results we have assumed that eventually the faster growing solution (the one proportional to $a^3$) will dominate the other one (the constant one).

The result  ~(\ref{epseq}) is our proposed modification of Maldacena's consistency relation, valid for the model of ultra slow-roll inflation. However, we have to admit that strictly speaking it is not a true consistency relation. The original consistency relation relates three observables: power spectrum, bispectrum and spectral index. In this case, the power spectrum and bispectrum are related via the factor $(2\eps+\eta)$, which is not a direct observable (although for ultra slow-roll it simply asymptotes to $-6$).

\setcounter{equation}{0}
\section{More general non-attractor backgrounds} \label{sec:general-non-attractor}

Before concluding, let us show how our arguments may be implemented in more general non-attractor backgrounds.  To this extent, let us consider inflation in the context of $P(X)$-models, where the Lagrangian is given by a general function of the inflaton field in the following form
\be
\mathcal L = P(X, \phi), \qquad X =  - \frac{1}{2} (\partial \phi)^2 .
\ee
We remind the reader that even in these more general models, we still restrict ourselves to models in which $\epsilon$ is zero or decaying very rapidly. The background equations of motion are found to be given by
\ba
(2 X P_{XX} + P_X) \ddot \phi + 2 X P_{X \phi} + 3 H P_X \dot \phi - P_{\phi} = 0,   \label{eq-motion-PX-1} \\
3 H^2 = 2 X P_X  - P ,  \label{eq-motion-PX-2}  
\ea
where $P_X \equiv \partial_X P$, $P_{\phi} \equiv \partial_\phi P$, and now $X = \dot \phi^2 / 2$. One can also deduce the additional convenient equation 
\be
\dot H = - X P_X, \label{eq-motion-PX-3} 
\ee
which in turn, tells us that 
\be
\epsilon = \frac{X P_X}{H^2}.   \label{epsilon-PX}
\ee
These equations of motion admit both attractor and non-attractor backgrounds. Of course, we will be interested in non-attractor backgrounds as long as they satisfy the condition $\epsilon \ll 1$ (in order to have a quasi-de Sitter geometry). Our next challenge is to identify a symmetry of the Lagrangian allowing us to deduce how the perturbations evolve after horizon crossing. A transformation of the Lagrangian under such an alleged symmetry will satisfy
\be
\Delta \mathcal L = P_X \Delta X + P_\phi \Delta \phi = 0 ,
\ee
from where one reads $P_\phi = - P_X  \Delta X / \Delta \phi$. Inserting this result back into the background equations of motions we find:
\be
(2 X P_{XX} + P_X) \ddot \phi \Delta \phi + 2 X P_{X \phi} \Delta \phi + 3 H \dot \phi P_X \Delta \phi+ P_{X}  \Delta X = 0 . \label{eq-with-symmetry}
\ee
By noticing that $\Delta X = \dot \phi \Delta \dot \phi = \dot \phi \, d (\Delta \phi) / dt$, eq.~(\ref{eq-with-symmetry}) may now be integrated once. The solution is found to be given by
\be
\Delta \phi = \frac{\Delta C}{a^3 \sqrt{2 X} P_X } , \label{PX-shift-symmetry}
\ee
where $\Delta C$ denotes an integration constant. Thus, a transformation of the inflaton field of the form $\phi \to \phi' = \phi + \Delta \phi$ with $\Delta \phi$ given by~(\ref{PX-shift-symmetry}) constitutes a symmetry of the background. Notice that in the specific case of ultra slow-roll, examined in Section~\ref{USR}, we have $P = X - V_0$, $P_X=1$, and $\sqrt{2 X} = \dot \phi \propto a^{-3}$, and eq.~(\ref{PX-shift-symmetry}) is reduced to a shift symmetry $\Delta \phi = \rm{constant}$, as it should. On the other hand, in the case of a slow-roll background, we recover $\Delta \phi \sim a^{-3}$, reminding us about the attractor nature of slow-roll inflation.

\subsection{Symmetry and long wavelength perturbations}

To explore the consequences of eq.~(\ref{PX-shift-symmetry}) on the perturbations, let us stick to the metric~(\ref{metric-flat}) written in flat gauge. Then, repeating the arguments of Section~\ref{sec:long-wavelengths-flat}, we see that $\delta N_\ell \propto \epsilon$. Then, if the non-attractor background is such that $\epsilon \to 0$ quickly, eq.~(\ref{PX-shift-symmetry}) informs us that $N_\ell \to 1$, and the value of the effective scalar field $\phi_{\rm eff}$ at a given patch of the universe centered at $\x$ will be given by:
\be
\phi_{\rm eff} (t , \x) = \phi(t) + \Q_\ell (t , \x) .
\ee
Then, because this form of the field has to be consistent with the symmetry underlying eq.~(\ref{PX-shift-symmetry}), we conclude that the long wavelength behavior of the scalar perturbation is given by:
\be
\Q_\ell = \frac{C}{a^3 \sqrt{2 X} P_X } .
\ee
This in turn tells us that there is a perturbation $\F$, proportional to $\Q$, that freezes outside the horizon, and is given by:
\be
\F \equiv a^3 \sqrt{2 X} P_X \Q .
\ee
On the other hand, let us recall that, up to second order in the fields, $\Q$ is related to the adiabatic curvature perturbations $\R$ in the following way
\be
\Q =  \frac{\dot \phi }{H} \left[ \R + \frac{\eta}{4 c_s^2} \R^2 + \frac{1}{H c_s^2} \R  \dot \R \right] + \cdots, \label{QR-cs}
\ee
where this time we have accounted for the presence of the speed of sound $c_s$ of curvature perturbations, which for the models examined up to this point  had a value $c_s = 1$. The explicit value of $c_s$ in terms of other background quantities is:
\be
c_s^2 \equiv \frac{P_X}{P_X + 2 X P_{XX}} .
\ee
Equation~(\ref{QR-cs}) further implies that:
\be
\F =  2 a^3 H \epsilon \left[ \R + \frac{\eta}{4 c_s^2} \R^2 + \frac{1}{H c_s^2} \R  \dot \R \right] + \cdots , \label{F-R}
\ee
where we have used the identity of eq.~(\ref{epsilon-PX}). Now, at linear order we have $\R =  \F / {2 a^3 H \epsilon }$. Therefore, in the long wavelength limit $\R_{\ell} \propto 1 / {2 a^3 H \epsilon}$, from where we deduce that at linear order
\be
\dot \R_\ell = - H (\eta + 3) \R_\ell ,
\ee
which is obtained after using the background equations of motion, and the fact that $\epsilon \ll 1$. Then, inserting this result back into eq.~(\ref{F-R}) we obtain
\be
\F_\ell =  2 a^3 H \epsilon  \left[ \R_\ell + \lambda \R_\ell^2  \right] , \label{F-R-2}
\ee
where
\be
\lambda = - \frac{3}{4 c_s^2}(4+ \eta) ,
\ee
Equation~(\ref{F-R-2}) is what we need to derive the squeezed limit for the bispectrum of $\R$. We examine this derivation in the following discussion.

\subsection{A general squeezed limit for non-attractor inflation} 

In what follows, we derive a general expression for the squeezed limit of non-Gaussianity, valid for general non-attractor models.
To start with, by using eq.~(\ref{3-point-f-g}) with $f \equiv  \F / {2 a^3 H \epsilon} $ and $g \equiv \R$, we obtain:
\be
 \langle \mathcal R (\x) \mathcal R (\y) \mathcal R (\z) \rangle  =  \frac{1}{8 a^9 H^3 \epsilon^3} \langle \F (\x) \F (\y) \F (\z) \rangle - 2 \lambda \left[  \langle \mathcal R (\x) \mathcal R (\y) Ê\rangle  \langle  \mathcal R (\x)  \mathcal R (\z) \rangle  + {\rm cyclic} \right] . \label{bispectrum-R-R}
\ee
Then, we may compute $\langle \F(\k_1) \F(\k_2) \F(\k_3) \rangle$ following the same procedure employed in Section~\ref{comp} for the case of ultra slow-roll. In this case, we obtain
\be
 \lim_{\k_3 \to 0} \langle \F(\k_1) \F(\k_2) \F(\k_3) \rangle  = (2 \pi)^3  \delta^3 (\k_1 + \k_2 + \k_3) P_{\F}( k_1) \left[ \frac{\partial}{\partial \F_\ell}  P^B_{\F}(k_3) \right]_0 ,
\ee
where $P^B_{\F}(k)$ represents the power spectrum for $\F$ computed in a background that is renormalized by the long wavelength contributions $\F_\ell$. This result may be reexpressed in terms of the curvature power spectrum as:
\be
 \lim_{\k_3 \to 0} \langle \F(\k_1) \F(\k_2) \F(\k_3)  \rangle  =  16 \, h_\F \, a^{12} H^4 \epsilon^4  (2 \pi)^3  \delta^3 (\k_1 + \k_2 + \k_3) P_{\R}( k_1) P_{\R}( k_3) . \label{bispectrum-F}
\ee
where we have defined:
\be
h_\F \equiv \left[ \frac{\partial}{\partial \F_\ell} \ln P^B_{\F}  \right]_0 .
\ee
Now, putting together eqs.~(\ref{bispectrum-F}) and (\ref{bispectrum-R-R}) in momentum space, we finally arrive to
\ba
 \lim_{\k_3 \to 0} \langle \R(\k_1) \R(\k_2) \R(\k_3) \rangle  &=&  2 \, h_\F \, a^{3} H \epsilon (2 \pi)^3  \delta^3 (\k_1 + \k_2 + \k_3) P_{\R}( k_1) P_{\R}( k_3)  \nn \\
&& - 4 \lambda ~(2\pi)^3\delta^3 (\k_1 + \k_2 + \k_3) P_{\R}(k_1)  P_{\R}(k_3) .
\ea
To simplify this expression, we notice that the first term of the right hand side should vanish in the long wavelength limit. Indeed, on the one hand the overall factor $2 a^{3} H \epsilon $ decreases quickly for non-attractor models (otherwise curvature perturbations would not grow). On the other hand, because $\F$ freezes in the long-wavelength limit, $P_{\F}$ must tend to a constant, independently of the value of $\F_\ell$ (as in the case of ultra slow-roll). These considerations lead to our final result valid for general non-attractor models:
\be
 \lim_{\k_3 \to 0} B_\R (\k_1, \k_2, \k_3) = \frac{3}{c_s^2}(4+ \eta)P_{\R}(k_1)  P_{\R}(k_3).
\ee
The overall factor at the right hand side is in general time-dependent, and therefore must be evaluated at the end of the non-attractor phase, which could coincide with the end of inflation, or could be the beginning of the attractor phase (see refs.~\cite{Namjoo:2012aa, Chen:2013aj} for a discussion on the phenomenological feasibility of this class of models). To finish, let us notice that our result reduces to the already known answer found in ref.~\cite{Chen:2013aj}, where a specific family of non-attractor solutions were studied.

\setcounter{equation}{0}
\section{Discussion and conclusions} \label{sec:conclusions}

We have studied the violation of the non-Gaussian consistency relation within single field non-attractor models of inflation, characterized by the fact that curvature perturbations do not freeze after horizon crossing. Our analysis was based purely on symmetry considerations: we found that the same arguments leading to the universality of the the non-Gaussian consistency relation for attractor models of inflation, may be reproduced to deduce a relation valid for non-attractor models. To achieve this, it was important to notice that while curvature perturbations $\R$ do not freeze after horizon crossing, other perturbations do ($\Q$ in the case of ultra slow-roll, and $\F$ in the case of more general non-attractor models). As a result, it is possible to further understand the violation of the consistency relation as a natural consequence of the super-horizon evolution of the modes, dictated by the background of the model, independently of the model being an attractor or not. Our results agree with those of previous analyses. In particular, we have re-derived the violation to the consistency relation in the context of ultra slow-roll inflation reported in ref.~\cite{Namjoo:2012aa}, and we have found a general expression for the squeezed limit of non-Gaussianity for non-attractor models which generalizes the results found in ref.~\cite{Chen:2013aj}. 

Our results offer a rationale to understand potentially large violations to the consistency relation, telling us how to relate the size of the violation to the evolution of the inflationary background. Given that future large scale structure surveys will offer substantially better constraints on non-Gaussianity than those currently available from CMB experiments, it is particularly timely to understand the difference between alternative mechanisms to violate the consistency relation. In this regard, one particularly interesting task ahead is to characterize the squeezed limit of other $n$-point correlation functions within non-attractor models. Just as in the case of single-field attractor models, it should be possible to derive specific relations among different $n$-point correlation functions, consistent with the symmetries of the inflationary background~\cite{Huang:2006eha, Li:2008gg, Seery:2008ax, Leblond:2010yq, Assassi:2012zq, Flauger:2013hra, Creminelli:2013mca}.

\subsection*{Acknowledgements}

We would like to thank Ana Ach\'ucarro, Vicente Atal, Diego Chialva, Jorge Nore\~na, Pablo Ortiz, Enrico Pajer and Yvette Welling for useful discussions and comments on the content of this work. This work was supported by the Fondecyt project number 1130777 (GAP), by the ``Anillo'' project ACT1122 funded by the ``Programa de Investigaci\'on Asociativa" (GAP \& SM) and by the Fondecyt 2015 Postdoctoral Grant 3150126 (SM).

\begin{appendix}

\renewcommand{\theequation}{\Alph{section}.\arabic{equation}}

\section{Evolution of modes in ultra slow-roll inflation} \label{app:usr}

In this appendix we explicitly compute the behavior of curvature perturbations for both standard attractor slow-roll inflation and for non-attractor ultra slow-roll inflation, to linear order. These computations are of course complementary to the derivations based on symmetry arguments found in Sections~\ref{attractor-long-wavelengths} and~\ref{USR}. However, we will use them to re-derive the freezing of the modes of $\Q$ on super-horizon scales

Let us start by quickly reviewing the freezing on super-horizon scales of curvature perturbations $\R$. In single-field models it is common lore to define the canonically normalized perturbation $v$ in terms of $\R$ as:
\be
v \equiv a \frac{\dot{\phi}}{H} \R = a\left(\frac{\dot{\phi}}{H} \psi + \delta \phi\right).  \label{defv}
\ee
In terms of its Fourier modes $v_k$ the equation of motion is
\be
v_k''-\frac{z''}{z}v_k + k^2 v_k=0, \qquad z\equiv \frac{a\dot{\phi}}{H},   \label{eqv}
\ee
where the prime $'$ denotes a derivative with respect to conformal time $\tau$. To zeroth order in the slow-roll parameters, in standard slow-roll inflation we find that $z''/z=2/\tau^2$, where conformal time may be written as $\tau = 1/(aH)$ where $H$ is constant. Then, in the long wavelength limit we obtain:
\be
v_k''-\frac{2}{\tau^2}v_k=0,\qquad \rightarrow \qquad v_k = c_1 \tau^2 + \frac{c_2}{\tau}  \propto \frac{c_1}{a^2} +c_2 a. \label{solv}
\ee
To zeroth order in the slow roll parameters the factor $\dot{\phi}/H$ in~(\ref{defv}) does not change in time, so we find that the modes $\R_k$ indeed freeze on super-horizon scales. A more detailed computation shows that this freezing is exact up to all orders in slow-roll parameters. For our discussion the rough sketch above suffices.

How does this situation change in ultra slow roll-inflation? In this case, $\eps = \dot{\phi}^2 / (2H^2)$ falls down as $a^{-6}$ and $\eta$ asymptotes to $-6$. Surprisingly, to zeroth order in $\eps$ and all orders in $\eta$ (compatible with the decoupling limit), we still find $z''/z=2/\tau^2$ and $\tau=1/(aH)$. To be precise, we still have
\be
\frac{z''}{z}=\frac{1}{\tau^2}\left[\nu^2-\frac{1}{4}\right], \qquad \nu^2=\frac{9}{4}\left(1+\frac{\eta}{3}\right)^2 \to \frac{9}{4},
\ee
as $\eta\to -6$. Therefore, the solution found in~(\ref{solv}) for the modes $v_k$ is still valid. Things change, however, in the conversion to $\R_k$, for which we now find the long wavelength limit:
\be
\R_k = \frac{1}{a} \frac{H}{\dot{\phi}} v_k \sim a^3.
\ee
As we already found in~(\ref{growmod}) by our symmetry argument, in ultra slow-roll inflation the modes $\R_k$ do not freeze on super-horizon scales.  However, from the above it is clear (at least up to zeroth order in $\eps$ and all orders in $\eta$), that the modes of a related perturbation
\be
\Q \equiv \frac{\dot{\phi}}{H}\psi +\delta \phi , \qquad \left(=\frac{\dot{\phi}}{H}\R\right)   \label{rel} 
\ee
do freeze on super-horizon scales.   Note that $\Q$ is as perfectly gauge invariant as $\R$, since the two only differ by a background quantity. Therefore, both variables share the same scalar spectral index $n_\R = n_\Q$. (The fact that the modes $\R_k$ keep evolving after horizon crossing does not take away that at any desired moment in time the relative differences in power between modes of different wavelengths are encoded in $n_\R$.)

\end{appendix}

\end{document}